# High-energy spin and charge excitations in electron-doped copper oxide superconductors

K. Ishii[1,*], M. Fujita[2], T. Sasaki[2], M. Minola[3], G. Dellea[3], C. Mazzoli[3], K. Kummer[4],
G. Ghiringhelli[3], L. Braicovich[3], T. Tohyama[5], K. Tsutsumi[2], K. Sato[2], R. Kajimoto[6],
K. Ikeuchi[7], K. Yamada[8], M. Yoshida[1,9], M. Kurooka[9], J. Mizuki[1,9]

[1]*SPring-8, Japan Atomic Energy Agency, Sayo, Hyogo 679-5148, Japan*
[2]*Institute for Materials Research, Tohoku University, Katahira, Sendai 980-8577, Japan*
[3]*CNR-SPIN and Dipartimento di Fisica, Politecnico di Milano, piazza Leonardo da Vinci 32, I-20133 Milano, Italy*
[4]*European Synchrotron Radiation Facility, 6 rue Jules Horowitz, F-38043 Grenoble, France*
[5]*Yukawa Institute for Theoretical Physics, Kyoto University, Kyoto 606-8502, Japan*
[6]*Materials and Life Science Division, J-PARC Center, Japan Atomic Energy Agency, Tokai, Ibaraki 319-1195, Japan*
[7]*Research Center for Neutron Science and Technology, Comprehensive Research Organization for Science and Society, Tokai, Ibaraki 319-1106, Japan*
[8]*Institute of Materials Structure Science, High Energy Accelerator Research Organization, Tsukuba, Ibaraki 305-0801 Japan*
[9]*School of Science and Technology, Kwansei Gakuin University, Sanda, Hyogo 669-1337, Japan*

**The evolution of electronic (spin and charge) excitations upon carrier doping is an extremely important issue in superconducting layered cuprates and the knowledge of its asymmetry between electron- and hole-dopings is still fragmentary. Here we combine x-ray and neutron inelastic scattering measurements to track the doping dependence of both spin and charge excitations in electron-doped materials. Copper $L_3$ resonant inelastic x-ray scattering spectra show that magnetic excitations shift to higher energy upon doping. Their dispersion becomes steeper near the magnetic zone center and deeply mix with charge excitations, indicating that electrons acquire a highly itinerant character in the doped metallic state. Moreover, above the magnetic excitations, an additional dispersing feature is observed near the $\Gamma$-point, and we ascribe it to particle-hole charge excitations. These properties are in stark contrast with the more localized spin-excitations (paramagnons) recently observed in hole-doped compounds even at high doping-levels.**

The differences and similarities between the electron- and hole-doped classes of superconductors have been one of the central subjects in the studies of the copper oxides [1]. They are crucially important for understanding the emergence of high-$T_c$ superconductivity—and likewise of Mott insulator behavior—in strongly correlated electron systems. Extensive works for this purpose have been done since the discovery of the high-$T_c$ superconductivity and it is now established that carrier doping to the parent antiferromagnetic Mott insulator induces a metallic state, where the superconductivity in the cuprates emerges. Electrons in the metallic state often reveal both itinerant and localized characters, and the dual nature is a difficulty of the doped Mott insulators. It is generally believed that superconductivity in cuprates is intimately related to the antiferromagnetic spin fluctuation and inelastic neutron scattering (INS) has been widely used for studying the spin dynamics in the reciprocal lattice space. However momentum-resolved excitation spectra of the charge degrees of freedom have been less explored because of the limitations of this experimental technique, even though both degrees of freedom should be clarified for the comprehensive understanding of cuprates.

Magnetic excitations of the undoped parent compounds are fairly well described by spin-wave theory and the superexchange interaction between the nearest-neighbor Cu spins is found to be of the order of 100 meV [2-4]. Upon hole doping, the magnetic excitations change substantially and the "hour-glass" dispersion appears [5,6]. Below the energy $E_{cross}$ at the waist of the hour-glass, the excitations disperse downward to the incommensurate wave vector away from the antiferromagnetic ordering vector $\mathbf{q}_{AF} = (0.5,0.5)$ of the parent compounds. On the other hand, the dispersion of the excitations above $E_{cross}$ remains rather similar to that of the antiferromagnetic undoped compounds and a magnon-like mode survives. Very recently, resonant inelastic x-ray scattering (RIXS) at the Cu $L_3$-edge has become an alternative to neutron scattering to measure momentum-resolved magnetic excitations [7,8]. RIXS complementarily covers the momentum space around the Γ point, where high-energy magnetic excitations are hard to measure with INS, but cannot reach $q_{AF}$. RIXS studies of the hole-doping dependence clarified that intense magnetic excitations remain as a "paramagnon" even in the overdoped region and its dispersion relation and energy-integrated spectral weight appear to be almost independent of the hole concentration [9-12]. It is argued from the result that the high-energy magnetic fluctuation is a vital source of the superconductivity [9].

While high transition temperatures ($T_c$) are realized in hole-doped superconductors, exceeding 90 K in some materials, the transition temperatures of electron-doped superconductors are limited below ~ 25 K. Magnetic excitations of the electron-doped cuprates differ from the hole-doped ones; they exhibit commensurate peaks centered at $q_{AF}$ at low energy [13]. Just a few INS works studying the high-energy magnetic excitations have been published and they have covered a restricted number of electron concentrations [14,15] and, due to the cross-section limitation, only in the vicinity of $q_{AF}$. This shortcoming can be overcome in part by using modern spallation neutron sources like J-PARC, but another complementary technique is required to cover the wide energy and momentum ranges. RIXS at the Cu $L_3$-edge perfectly matches these requirements.

Here we use Cu $L_3$ edge RIXS and its complements, INS and Cu K-edge RIXS, for studying the electron-doping dependence of high-energy spin excitations and their counterpart, charge excitations, comprehensively. We chose $Nd_{2-x}Ce_xCuO_4$ (NCCO) and $Pr_{1.40-x}La_{0.60}Ce_xCuO_4$ (PLCCO) for RIXS and INS measurements, respectively. This is because the Pr $M_5$-edge is too close to the Cu $L_3$-edge to perform RIXS experiment and the large magnetic moment of Nd hampers the measurement of magnetic INS. We will later show that magnon dispersions of the undoped materials (x = 0) are very similar to each other, which is why the NCCO data obtained with RIXS and the PLCCO INS data can be compared meaningfully. Figure 1(a) shows a schematic phase diagram of $(Nd,Pr,La)_{2-x}Ce_xCuO_4$ as a function of the Ce concentration. We selected representative Ce concentrations for the present study: x = 0 (undoped antiferromagnet), 0.075 (doped antiferromagnet), 0.15 (optimally-doped superconductor), and 0.18 (overdoped superconductor) for NCCO and x = 0, x = 0.08, and 0.18 for PLCCO. The temperature of the samples was respectively kept at 20 K and 6 K for RIXS and INS, which are below the antiferromagnetic or superconducting transition temperature.

**Results**

We start with Cu $L_3$-edge RIXS. Because of the high two-dimensionality of the electronic properties, we discuss the momentum in the CuO$_2$ plane ($\mathbf{q}_{\parallel}$) in the tetragonal notation ($a = b \sim$ 3.945 Å, c ~ 12.1 Å [16]) by neglecting the variation of the momentum along the c$^*$-axis. The experimental configuration ($q_{\parallel} > 0$ and $\pi$ incident polarization), schematically presented in Fig. 1(b), is chosen to give the largest spectral weight of the spin-flip transition within the d$_{x2-y2}$ orbital as compared to the other channels (charge and lattice excitations) [7]. Experimental details are described in the Methods. Figure 1(c) shows the doping dependence at $\mathbf{q}_{\parallel}$ = (0.23,0). As demonstrated in Ref. [8], the low-energy spectrum of the undoped compound (x = 0) can be univocally decomposed into the elastic (purple), single-magnon (red), and multi-magnon (cyan) components, and the single-magnon component is dominant. Upon electron doping, the spectral weight clearly moves to higher energy and the width of the peak broadens. Moreover, as typically shown for x = 0.15 in Fig. 1(d), we found an additional peak (green) near the Γ-point in the doped compounds. It probably has charge origin, as discussed later.

In Figure 2, we summarize the momentum and doping dependence of the Cu L$_3$-edge RIXS spectra of NCCO. The results for x = 0 [Fig. 2(a)] are similar to those in the previous studies on undoped cuprates [4,8,9]. The spectra are decomposed into three components as in the Fig. 1(c), and the elastic and spin-flip components are found to be resolution-limited. The observed momentum dependence of the single-magnon components follows the dispersion of the spin-wave of the antiferromagnetic square lattice.

The spectra of the doped compounds in Figs. 2(b)-(e) cannot be decomposed in the same way as the undoped ones in Fig. 2(a) because of the intrinsic broadening of the peaks. We analyze the spectra of the doped compounds for large $\mathbf{q}_{\parallel}$ by a single inelastic component, which has been adopted from the published works of the hole-doped compounds. The main portion of the spectral weight is assigned to the spin-flip excitations [9-12]. This assignment is supported by a recent theoretical study which demonstrates that the RIXS spectra of the $\pi$-incident polarization agree well with the dynamical spin correlation function, $S(\mathbf{q},\omega)$, even without outgoing polarization analysis [17]. In addition, the two-magnon component observed by Raman scattering is weakened by electron doping [18], and it is therefore acceptable to neglect the two-magnon component in the analysis of the doped compounds. As inelastic component, we use an antisymmetrized Lorentzian function convoluted by the Gaussian resolution. The spectra for $h \geq 0.14$ for $\mathbf{q}_{\parallel} = (h,0)$ and $h \geq 0.10$ for $\mathbf{q}_{\parallel} = (h,h)$ can be fitted well by the simple model of the resolution-limited elastic component and the single inelastic peak. The peak position and width obtained from the fittings are plotted against the momentum in Figs. 3(a) and 3(b), respectively. We also plot the single-magnon components of the undoped compounds. Our RIXS results clearly demonstrate that the magnetic excitations shift to higher energy and broaden upon electron doping. It is notable that the high-energy shift of the magnetic excitation upon electron doping is consistent with a recent theoretical calculation [17].

While the electron-doping dependence in a large portion of the momentum space around the Γ-point is unveiled by the Cu $L_3$-edge RIXS, INS can cover the momentum space near $q_{AF}$. Figures 4(a)-(c) show magnetic INS spectra of PLCCO. For x = 0 the spin-wave dispersion, indicated by the open circles in Fig. 4(a), closely traces that of NCCO (x = 0) as obtained by RIXS: in Fig. 3(a) the INS data of undoped PLCCO is represented by the black crosses. When electrons are doped, the spectral weight at 100-300 meV accumulates gradually at $q_{AF}$ and it stands steeply at $q_\parallel = q_{AF}$. Such a doping effect is clearer in the constant energy cut in Fig. 4(e), where INS intensity at 280 ± 20 meV is plotted as a function of momentum transfer. While the peak appears near the Brillouin zone boundary (ZB) in x = 0, it remains at the magnetic zone center (ZC) in x = 0.18. In other words, the spectral weight at a certain momentum away from $q_{AF}$ moves toward higher energy upon doping, consistently with the Cu $L_3$-edge RIXS results around the Γ point. A detailed analysis of the INS results on PLCCO will be published elsewhere. Such spin excitations have been reported in electron-doped $Pr_{1-x}La Ce_x CuO_4$ below 180 meV [15,19], and the present study confirmed that they extend up to 300 meV. Alternatively, an analysis based on a two-dimensional spin-wave dispersion gave a larger nearest-neighbor coupling for $Pr_{1-x}La Ce_x CuO_4$ (x = 0.12) than that for $Pr_2 CuO_4$ [14]. It is noted that low-energy excitations (< 15 meV) behave oppositely; the magnetic peak centered at $q_{AF}$ becomes broader in the momentum direction upon electron doping [20,21]. In Fig. 3(c), we schematically summarize the magnetic excitation dispersion obtained by RIXS and INS.

Now we turn to the additional peak [green in Fig. 1(d)] above the magnetic excitation observed with Cu $L_3$-edge RIXS. In Fig. 3(d), we plot the peak positions of NCCO (x = 0.075, 0.15 and 0.18) obtain by the fitting analysis of the Cu $L_3$-edge RIXS [green lines in Fig. 2(b)-(e)] on the intensity map of the Cu K-edge RIXS re-plotted from Ref. [22]. As spin-flip transitions are forbidden in K-edge RIXS, the intensity around 1-2 eV can be assigned only to charge excitations. Thus the dispersive features with momentum-dependent width are ascribed to intraband charge excitations (incoherent particle-hole excitations) within the upper Hubbard band, while interband excitations across the charge transfer gap are concentrated around 2 eV near the zone center [22,23]. The peak positions in the Cu $L_3$-edge RIXS smoothly connect to the dispersive features in the Cu K-edge RIXS, and the peak width becomes broader with increasing $|q_\parallel|$ in the same way as in the K-edge spectra. In Fig. 3(e), we show a K-edge RIXS spectrum of NCCO (x = 0.18), which was measured recently with better energy resolution than that in Ref. [22]. Both intraband and interband transitions are observed on the tail of an excitation at higher energy, and the peak position of the former agrees with that at $q_\parallel$ = (0.09,0) in the Cu $L_3$-edge RIXS of the same compound. Therefore, it seems reasonable to ascribe the additional excitation in the Cu $L_3$-edge RIXS to the same charge origin as the dispersive mode in the K-edge RIXS, and it is possibly particle-hole excitations within the upper Hubbard band. The existence of the particle-hole excitations has been established in the *t-J* or Hubbard models [See Supplementary Information]. The energy of particle-hole excitation is of the order of the transfer energy (*t* ~ 400 meV), and it should be higher than the magnetic excitations at the energy scale of the exchange interaction (*J* ~ 100 meV) in the models, which agrees with our assignment. Plasmon excitations might be another possible origin of the additional excitation in the Cu $L_3$-edge RIXS. A theoretical work proposed that RIXS can probe dispersive plasmon excitations whose energy gap at the Γ point depends on the out-of-plane momentum ($q_z$) [24]. However solid assignment of the plasmon excitations in the experimental RIXS spectra has not been done so far. It may be tested by detailed dependence on $q_z$.

The doping evolution of the charge excitations is momentum-dependent. At the low-|$\mathbf{q}_\parallel$| region, the spectral weight of the charge excitations shifts to higher energy with increasing electron concentration, which is evidenced by the Cu L-edge RIXS in Fig. 3(d). On the other hand, the spectral weight increases almost at the same energy with doping at high |$\mathbf{q}_\parallel$| and high energy, for example, $h > 0.2$ for $\mathbf{q}_\parallel = (h,0)$ and above 1 eV. The latter has been confirmed by the K-edge RIXS [22,25]. This doping evolution is reproduced in the theoretical calculation of the dynamical charge correlation function, $N(\mathbf{q},\omega)$ [See Supplementary Fig. S2]. At the smallest **q** defined in the system of the calculation [**q** = (0.2,0.1)], the spectral weight below ω ~ *t* is larger than that in high-energy region near ω ~ 2*t* for x = 0.1. Note that the low-energy weight is governed by the charge motion scaled by *J* due to spin-charge coupling, while the high-energy weight is caused by an incoherent motion of charge scaled by *t* [26]. With increasing x, the low-energy weight is suppressed accompanied with the increase of the incoherent weight, leading to the shift of peak position to higher-energy region. Such a shift is consistent with the observed data shown in Fig. 3(d). For larger value of **q**, the incoherent charge motion dominates the spectral weight independent of x.

**Discussion**

From the Cu $L_3$-edge RIXS, complemented by INS and Cu K-edge RIXS, we have thus clarified the doping evolution of the spin and charge excitations in the electron-doped copper oxide superconductors. Notably, the spin-wave magnetic excitations of the undoped insulator shift to higher energy upon electron doping over a wide momentum space. This is in distinct contrast to the hole-doped case, where spectral distribution of the magnetic excitations broadens but keeps its energy position almost unchanged upon doping, namely, it follows rather closely the spin-waves dispersion of the parent compound. It means that, in hole-doped copper oxides, the spin excitation at high energy is a remnant mode of the parent antiferromagnetic insulator and the spin dynamics has localized nature. On the other hand, the spectra reported in the present study reflect the strongly itinerant character of the electron-doped copper oxides. Actually, it was pointed out in Ref. [15] that the high-energy excitations around $\mathbf{q}_{AF}$ in INS of PLCCO bear a close similarity of the spin excitations in the nearly antiferromagnetic metals, vanadium-doped Cr [27] and iron-doped $Mn_3Si$ [28].

The itinerant character of electron-doped compounds can be observed also by exploiting the polarization dependence of Cu $L_3$-edge RIXS. It is known that at large positive values of $q_\parallel$, spin-flip magnetic excitations dominate in the π-polarization condition while they are negligible in the σ-polarization condition. At δ = 45°, the ratios of spin-flip ($d_{x2-y2}\downarrow \rightarrow d_{x2-y2}\uparrow$) to spin-conserved ($d_{x2-y2}\downarrow \rightarrow d_{x2-y2}\downarrow$) cross sections in the $Cu^{2+}$ single-ion model are 2.14 and 0.03 for π- and σ-polarizations, respectively [29]. In Fig. 2(f)-(i), we systematically compare the RIXS spectra measured with π- and σ-polarizations of incident photons. In the undoped compound [Fig. 2(f)], the single-magnon is clearly enhanced with π-polarization, as previously observed in other parent antiferromagnetic compounds [9,30], and multi-magnon excitations make a dominant contribution to the σ-polarization spectrum. Such single-magnon dominance in the polarization dependence seems to persist for x = 0.075, although to a lesser extent with respect to the hole-doped case [9]. When further increasing the electron concentration, the spectral shape becomes similar between the two polarizations; more precisely, the spectrum of the π-polarization is overlaid with that of the σ-polarization with its peak center located at slightly higher energy. This means that in optimal and over-doped compounds magnetic (spin-flip) excitations mix with the charge (spin-conserved) excitations in the same energy scale. Probably, the σ-polarization spectra are mainly given by a continuum of electron-hole pair excitations and, in the same energy range, the spin-flip channel can be considered as sort of Stoner excitations: the almost flat dispersion far from $\mathbf{q}_\parallel$ = (0,0) and (π,π) may be an indication of the particle-hole-like nature of magnetic excitations. Therefore the electron dynamics of the electron-doped cuprates has highly itinerant character in the sub-eV energy scale, and the concept of "paramagnon", which is valid up to the overdoped region in the hole-doping case [11,12], does not work any more.

**Methods**

**Sample preparation**

We prepared $Nd_{2-x}Ce_xCuO_4$ (NCCO) for resonant inelastic x-ray scattering (RIXS) and $Pr_{1.40-x}La_{0.60}Ce_xCuO_4$ (PLCCO) for inelastic neutron scattering (INS). Single crystals of NCCO (x = 0, 0.075, 0.15, and 0.18) and PLCCO (x = 0, 0.08, and 0.18) were grown by the traveling-solvent floating-zone method. Doped NCCO crystals were annealed at 900°C for 10 hours in the Ar gas atmosphere. The Néel temperatures determined by elastic neutron scattering measurements for NCCO (x = 0, 0.075, 0.15, and 0.18) and PLCCO (x = 0, 0.08, and 0.18) were [269(5), 218(5), 63(18) and 10(5) K] and [278(8), 228(5) and 155(18) K]. Superconducting transition temperatures of NCCO (x = 0.15 and 0.18) were [25(1) and 22(1) K].

**Cu L₃-edge RIXS**

The Cu L-edge RIXS experiments were performed at beam line ID08 of the ESRF using the AXES spectrometer with linearly polarized x rays, either perpendicular (σ) or parallel (π) to the scattering plane. Total energy resolution was 250 meV. The NCCO crystals were cleaved in the air and mounted on the spectrometer. The surface of the crystal was normal to the c axis, which was kept in the scattering plane so as to be scanned in the reciprocal lattice space spanned by either the [100]-[001] or the [110]-[001] axes. Temperature of the sample was kept at 20 K.

The experimental configuration is schematically presented in Fig. 1(b) of the main text. Momentum in the CuO$_2$ plane ($\mathbf{q}_\|$) is changed by rotating the sample. In the present study, we scanned in the range of $0° \leq |\delta| \leq 45°$ and it corresponds to the reciprocal lattice space of $0 \leq |h| \leq 0.38$ and $0 \leq |h| \leq 0.27$ for $\mathbf{q}_\| = (h,0)$ and $\mathbf{q}_\| = (h,h)$, respectively.

Supplementary Figure S1(a) shows x-ray absorption spectra of NCCO near the Cu L$_3$-edge. Incident photon for the RIXS measurements was tuned at the peak energy indicated by the arrow. In Fig. S1(b), we present RIXS spectra taken at the in-plane momentum $\mathbf{q}_\| = (0.23,0)$. Huge *dd* excitations (transition between the *d* orbitals) are observed at 1-3 eV region. Spectral shape of the *dd* excitations varies with the incident photon polarization and the experimental geometry, and it is characteristic to the four-fold coordination of Cu like CaCuO$_2$ [29]. Spectral weight below 1 eV comes from spin and charge excitations, which are the main subject in this paper.

We fit the spectra in the Figures 1(c,d) and 2 as follows. The elastic peak (purple line in the figures) is fit with a resolution-limited Gaussian function. The component of the spin-flip magnetic excitations (red) is modeled by an antisymmetrized Lorentzian function multiplied by the Bose factor and it is convoluted by the Gaussian resolution function. The width of the Lorentzian function is infinitesimal for the spectra of the undoped compound (x = 0) while it is variable for the doped compounds. The multi-magnon (cyan) and charge excitations (green) are assumed by a Gaussian function with variable width. We also add a smooth background (grey) from the tail of the *dd* excitations in the analysis. Because the decomposition of the elastic and magnetic components is difficult at very low $|\mathbf{q}_\||$ [$h \leq 0.05$ for $\mathbf{q}_\| = (h,0)$ and $h \leq 0.03$ for $\mathbf{q}_\| = (h,h)$] of the doped compounds, we do not include the elastic component for the analysis. Therefore we rely on INS for the peak positions of the magnetic component near the zone center.

**Cu K-edge RIXS**

The Cu K-edge RIXS experiments of NCCO (x = 0.18) were carried out at BL11XU of SPring-8. A channel-cut Si(333) monochromator and a Ge(733) analyzer were used and total energy resolution was 250 meV. The surface of the crystal was normal to the c-axis and the a- and c-axes were kept in the horizontal scattering plane. Temperature of the sample was 20 K.

**Inelastic Neutron Scattering**

INS spectra were measured by a direct-geometry chopper spectrometer 4SEASONS installed in the Materials and Life Science Experimental Facility in the Japan Proton Accelerator Research Complex. To cover a spin excitation in a wide energy range, the neutron incident energy of 357 meV was selected (half-width at half-maximum energy resolution of ~20 meV at the elastic position). The crystals were co-aligned so that the crystallographic c-axis is parallel to the incident neutron beam. The total weight of PLCCO with 0, 0.08 and 0.18 was 42.2, 42.6 and 28.1 grams, respectively. For all samples, the temperature was controlled to be 6 K by a closed cycle refrigerator.

Acknowledgements

We would like to acknowledge K. Tsutsui, E. Kaneshita, S. Wakimoto, and H. Kimura for invaluable discussion. We also thank H. Kimura for sample preparation. This work is financially supported by JSPS KAKENHI Grant Number 25400333, 23340093 and 22340097. The Cu $L_3$ RIXS experiments were performed at the beam line ID08 of the European Synchrotron Radiation Facility (ESRF) using the AXES spectrometer and we acknowledge the ESRF for provision of synchrotron radiation facilities. The INS experiments at J-PARC were carried out under Project No. 2012P0201. The Cu K-edge RIXS experiments at SPring-8 were performed with the approval of the Japan Synchrotron Radiation Research Institute (JASRI) (Proposals No. 2012B3502).

Author contribution

K. Is., M.F., and T.T. managed the project. M.F., T.S., K.T., K.S., and K.Y. grew the sample. K. Is., M.F., T.S., M.M., G.D., C.M., K.K., G.G. and L.B. performed the Cu L-edge RIXS experiment. M.F., T.S., K.T., K.S., K.Y. carried out the INS experiment assisted by R. K. and K. Ik. K. Is., Y.M., M.K., and J.M. conducted the Cu K-edge RIXS experiment. K. Is. and M.F. analyzed the experimental data. T.T. performed the theoretical calculation. K. Is., M. F. and T.T. wrote the paper.

Additional information

The authors declare no competing financial interests.

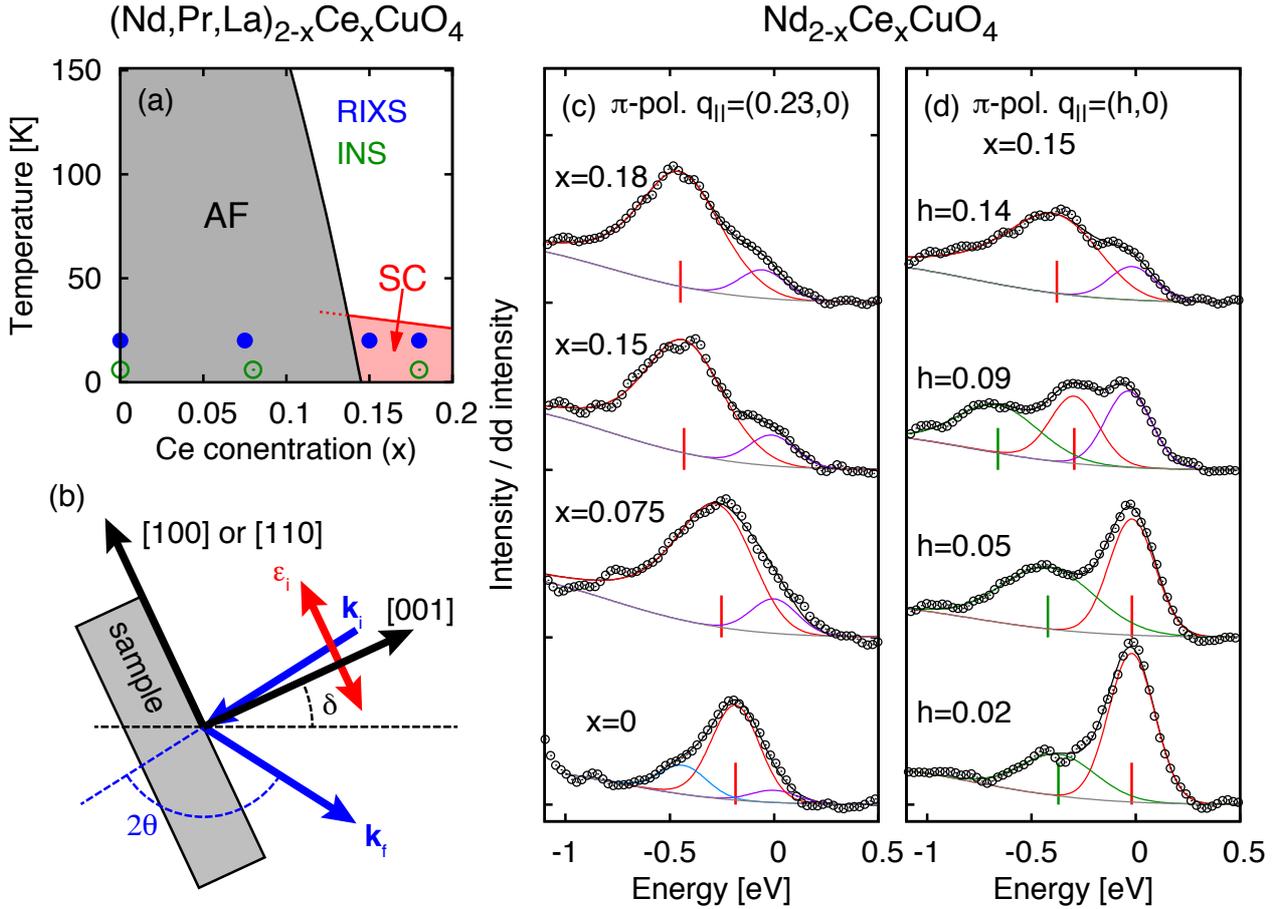

**Figure 1 | Schematic phase diagram of electron-doped copper oxide superconductors, experimental configuration, and typical Cu $L_3$-edge RIXS spectra. a**, Schematic phase diagram of the copper oxide superconductor $(Nd,La,Pr)_{2-x}Ce_xCuO_4$. AF and SC denote antiferromagnetic and superconducting phases, respectively. Filled and open circles indicate measured points of Cu $L_3$-edge RIXS and INS, respectively. **b**, Experimental configuration of Cu $L_3$-edge RIXS. The scattering angle ($2\theta$) is fixed at 130°. $\delta$ is the rotation angle of the sample from specular toward grazing emission conditions. We take $q_\parallel > 0$ [$h > 0$ in $\mathbf{q}_\parallel = (h,0)$ or $(h,h)$] for $\delta > 0$. **c,d**, Cu $L_3$-edge RIXS spectra of $Nd_{2-x}Ce_xCuO_4$ in the $\pi$-polarization conditions for (**c**) doping dependence at the fixed in-plane momentum $\mathbf{q}_\parallel = (0.23,0)$ and (**d**) momentum dependence of x = 0.15 near the $\Gamma$-point. The experimental spectra (open circles) are decomposed into a few components (solid lines). Peak positions of the spin-flip magnetic component (red) and charge excitations (green) are indicated by vertical bars.

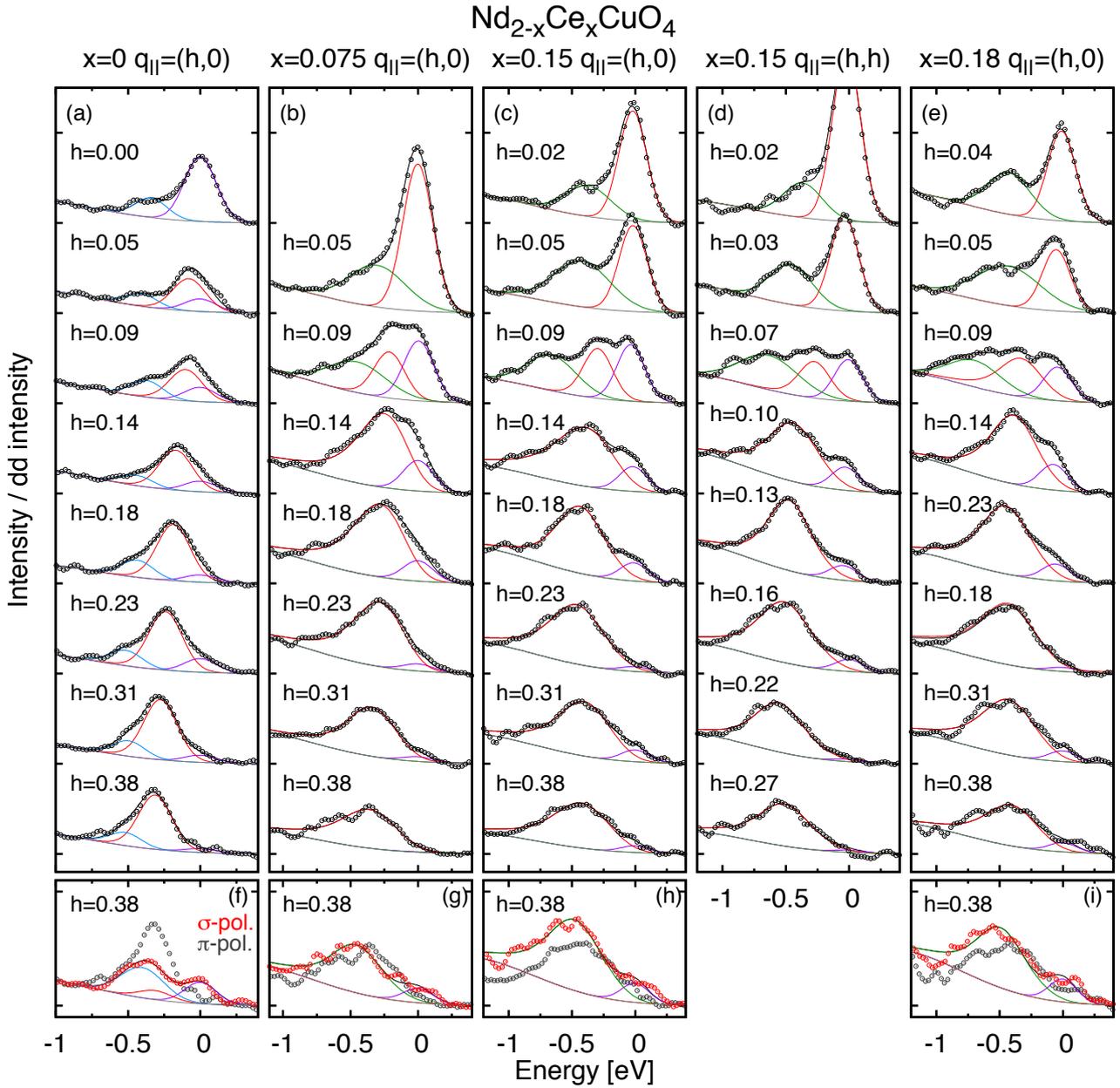

Figure 2 | **Experimental data of Cu $L_3$-edge RIXS.** (**a-e**) Momentum dependence of the RIXS spectra of $Nd_{2-x}Ce_xCuO_4$ measured in π-polarization condition. Open circles are the experimental data and solid lines are the fitting results of elastic (purple), single-magnon (red), multi-magnon (cyan), background from the tail of excitations at higher energy (grey) for the undoped compound, and sum of the all components (black). (**a**). In the doped compounds (**b-e**), we analyze the spectra using a single magnetic component (red) and an additional peak (green) at low $q_{||}$. Detailed fitting procedure is described in the Methods. (**f-i**) Comparison of the spectra observed with incident σ and π polarization at δ = 45°. Solid lines are the fitting results for σ-polarization.

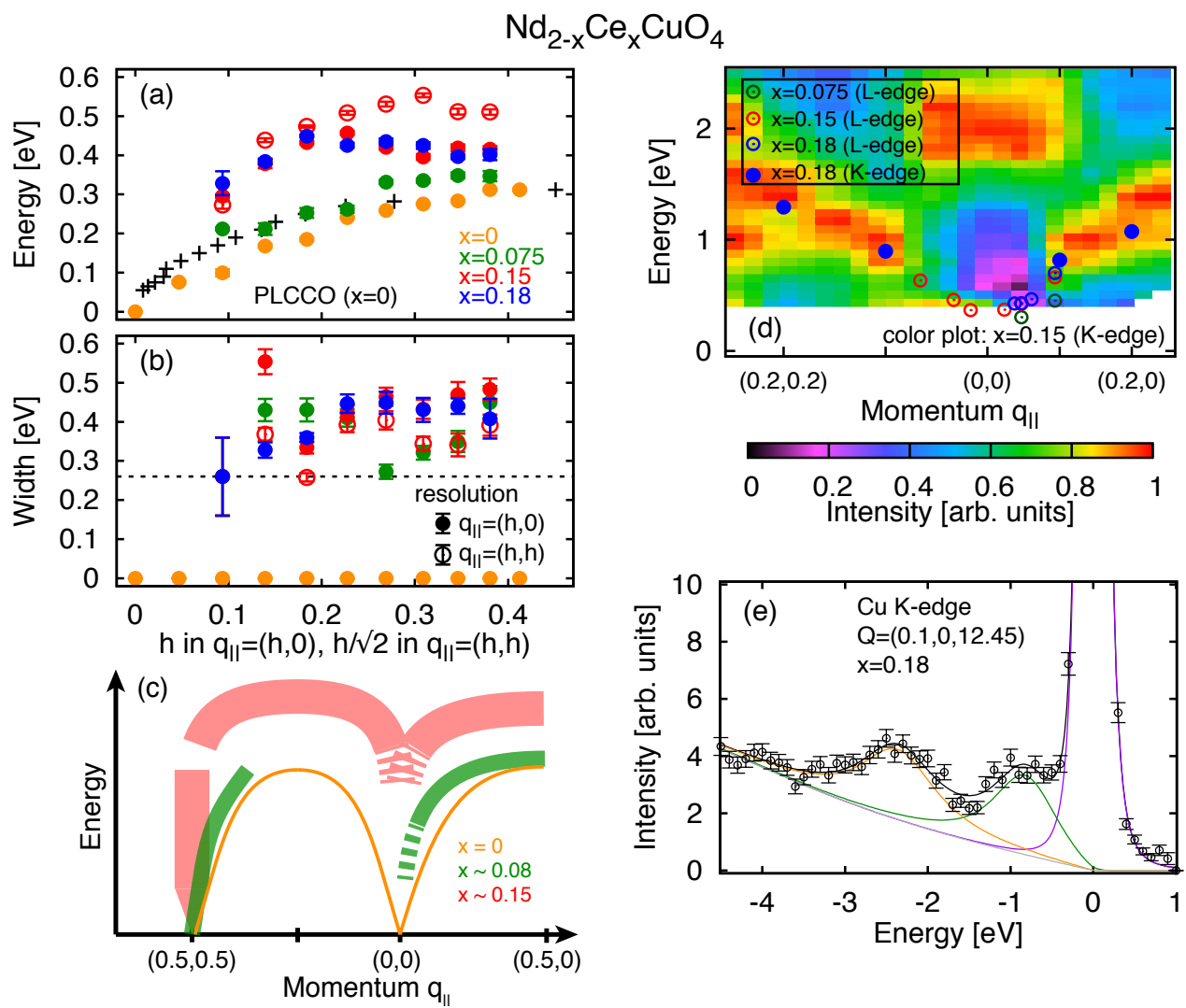

**Figure 3 | Peak position and width of the magnetic excitations, schematic of magnetic excitations in the energy-momentum space, and comparison with the Cu K-edge RIXS. (a,b)**, Peak position and width of magnetic excitations. We plot the single-magnon component for the undoped compound (x = 0). Peak positions of INS of undoped PLCCO are also shown by the crosses. The peak width of undoped NCCO (x=0) is fixed to infinitesimal in the analysis. **c**, Schematic of magnetic excitation in the energy-momentum space. **d**, Peak positions of the additional peak feature above the magnetic excitations observed with Cu $L_3$-edge (open circles). The background color plot is the intensity map of the K-edge RIXS which is re-plotted from the data in Ref. [19]. We also plot the peak position of the K-edge RIXS for x = 0.18 (filled circles). **e**, Cu K-edge RIXS spectrum of NCCO (x = 0.18). The spectrum is decomposed into elastic (purple), intraband (green) and interband (yellow) excitations, and the tail of high-energy excitation (grey).

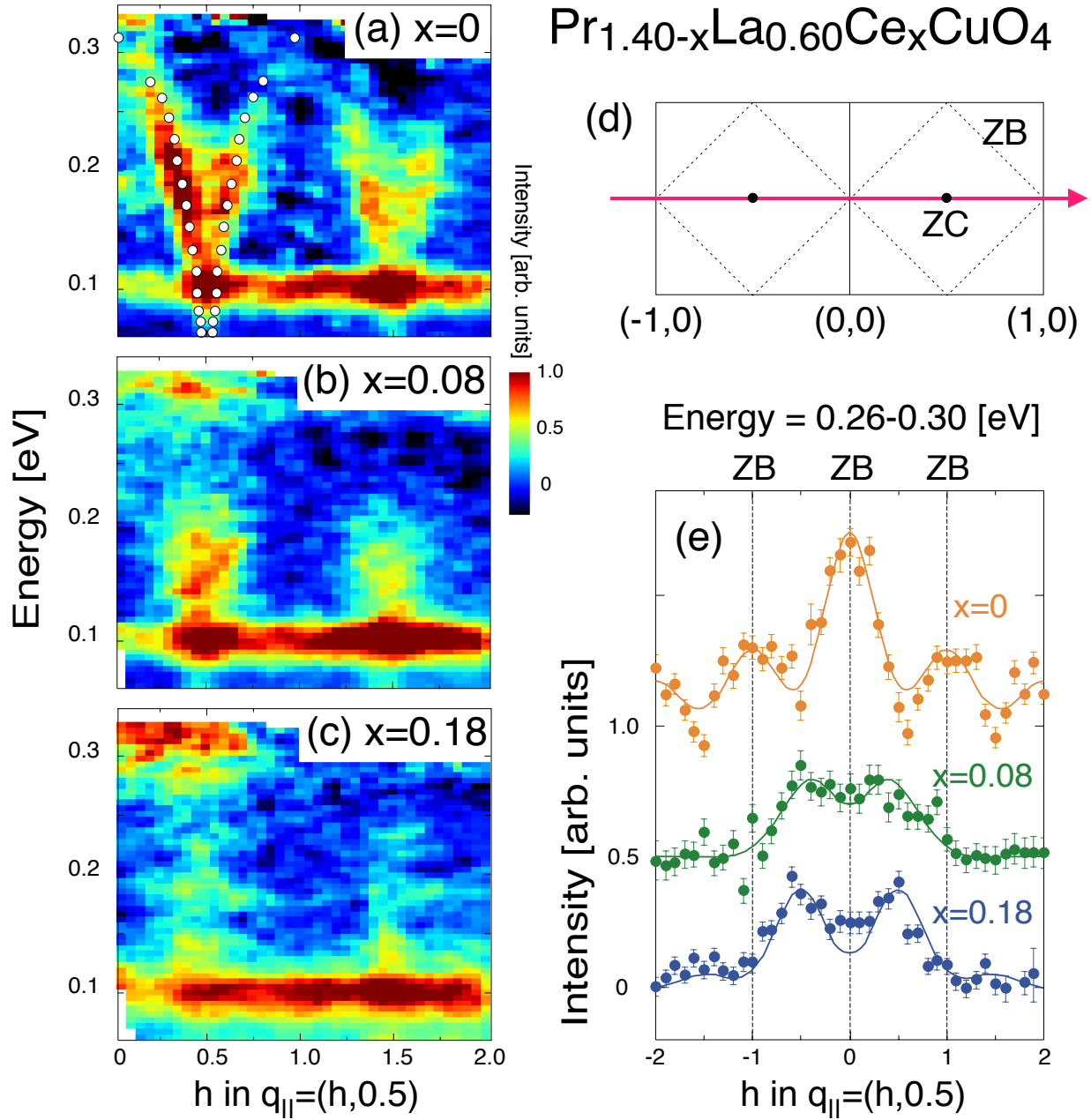

**Figure 4 | Magnetic INS intensity maps of PLCCO and their constant-energy cut.** (**a-c**), The momentum is $\mathbf{q}_{\parallel} = (h, 0.5)$. The neutron scattering intensity normalized to the sample volume after subtracting the non-magnetic background at (0.8, 0.2) is multiplied by $\omega^2$ to exhibit the entire spectrum with emphasizing the high-energy intensity. Open circles in (a) are the peak positions of magnon excitations. Dispersionless crystal field excitations of Pr are overlapped around 100, 320 meV. (**d**) Reciprocal lattice of the CuO$_2$ plane. ZC and ZB are the magnetic zone center and zone boundary, respectively. The abscissa of the figures (**a-c**, **e**) follows the arrow. (**e**) Constant-energy cut of INS intensity maps. Data are shifted vertically for clarity. To see the peak-shift from ZB to ZC upon doping, each data set was fitted by Gaussian functions. The solid lines are the best fitted result.

**Supplementary Information for "High-energy spin and charge excitations in electron-doped copper oxide superconductors"**

K. Ishii[1,*], M. Fujita[2], T. Sasaki[2], M. Minola[3,a], G. Dellea[3], C. Mazzoli[3], K. Kummer[4], G. Ghiringhelli[3], L. Braicovich[3], T. Tohyama[5], K. Tsutsumi[2], K. Sato[2], R. Kajimoto[6], K. Ikeuchi[7], K. Yamada[8], M. Yoshida[1,9], M. Kurooka[9], J. Mizuki[1,9]

*[1]SPring-8, Japan Atomic Energy Agency, Sayo, Hyogo 679-5148, Japan*
*[2]Institute for Materials Research, Tohoku University, Katahira, Sendai 980-8577, Japan*
*[3]CNR-SPIN and Dipartimento di Fisica, Politecnico di Milano, piazza Leonardo da Vinci 32, I-20133 Milano, Italy*
*[4]European Synchrotron Radiation Facility, 6 rue Jules Horowitz, F-38043 Grenoble, France*
*[5]Yukawa Institute for Theoretical Physics, Kyoto University, Kyoto 606-8502, Japan*
*[6]Materials and Life Science Division, J-PARC Center, Japan Atomic Energy Agency, Tokai, Ibaraki 319-1195, Japan*
*[7]Research Center for Neutron Science and Technology, Comprehensive Research Organization for Science and Society, Tokai, Ibaraki 319-1106, Japan*
*[8]Institute of Materials Structure Science, High Energy Accelerator Research Organization, Tsukuba, Ibaraki 305-0801 Japan*
*[9]School of Science and Technology, Kwansei Gakuin University, Sanda, Hyogo 669-1337, Japan*

**Supplementary figure**

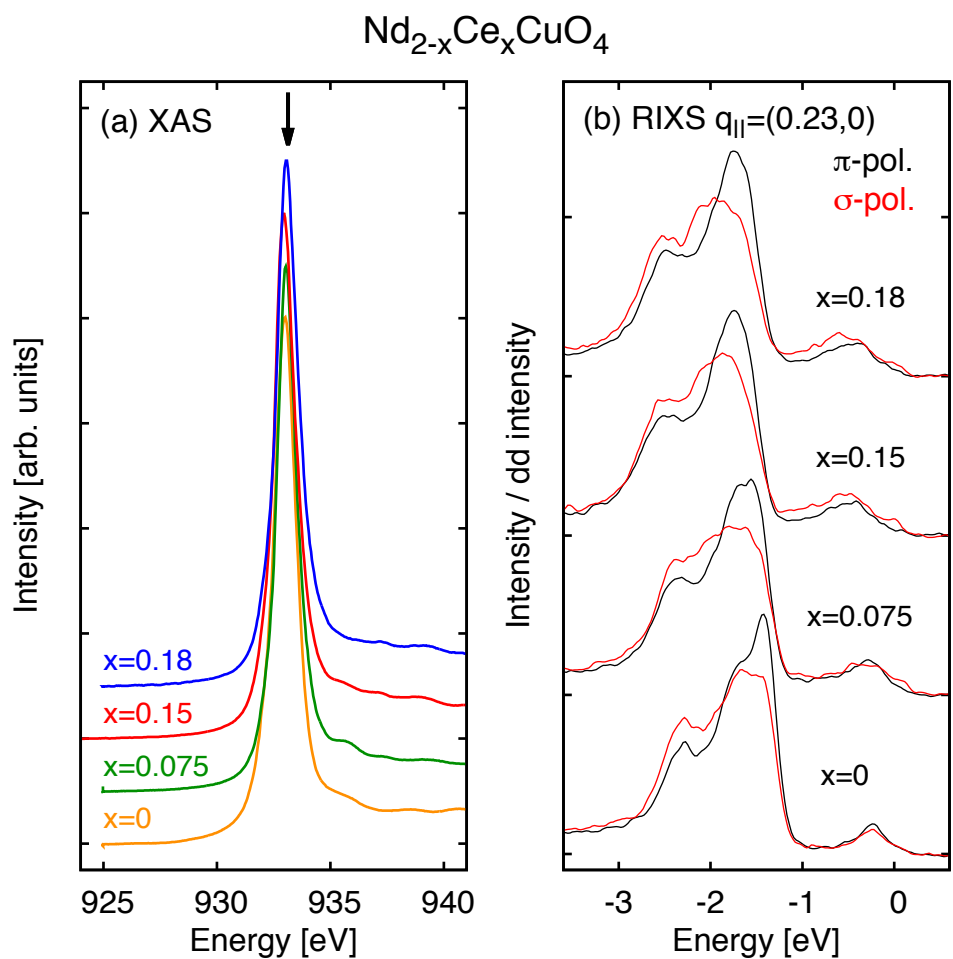

**Figure S1 | X-ray absorption and resonant inelastic x-ray scattering spectra of $Nd_{2-x}Ce_xCuO_4$ (NCCO) at the Cu $L_3$-edge. a**, X-ray absorption spectra of NCCO. Photon polarization is parallel to the *ab*-plane. The arrow indicates the incident photon energy for RIXS measurements. **b**, RIXS spectra of NCCO in a few eV region.

**Supplementary theory**

The dynamical charge correlation function was calculated for the *t-t'-t''-J* model with three-site term. The Hamiltonian is given by

$$H = H_{tt't''J} + H_{3s}$$

with

$$H_{tt't''J} = J \sum_{\langle i,j \rangle_{1st}} \mathbf{S}_i \cdot \mathbf{S}_j + t \sum_{\langle i,j \rangle_{1st},\sigma} \tilde{c}^\dagger_{i,\sigma} \tilde{c}_{j,\sigma} + t' \sum_{\langle i,j \rangle_{2nd},\sigma} \tilde{c}^\dagger_{i,\sigma} \tilde{c}_{j,\sigma} + t'' \sum_{\langle i,j \rangle_{3rd},\sigma} \tilde{c}^\dagger_{i,\sigma} \tilde{c}_{j,\sigma} + \text{H.c.}$$

and

$$H_{3s} = -\frac{J}{4} \sum_{\langle i,j \rangle_{1st},\langle i,j' \rangle_{1st}, j \neq j',\sigma} \left( \tilde{c}^\dagger_{j',\sigma} \tilde{n}_{i,-\sigma} \tilde{c}_{j,\sigma} - \tilde{c}^\dagger_{j',\sigma} \tilde{c}^\dagger_{i,-\sigma} \tilde{c}_{i,\sigma} \tilde{c}_{j,-\sigma} + \text{H.c.} \right),$$

where the summations $\langle i,j \rangle_{1st}$, $\langle i,j \rangle_{2nd}$, and $\langle i,j \rangle_{3rd}$ run over first, second, and third nearest-neighbor pairs, respectively. The operator $\tilde{c}_{i,\sigma} = c_{i,\sigma}(1 - c^\dagger_{i,-\sigma} c_{i,-\sigma})$ annihilates a localized particle with spin $\sigma$ at site $i$ with the constraint of no double occupancy, and $\mathbf{S}_i$ is the spin operator at site $i$. In the model, we set $t=1$, $t'=-0.25$, and $t''=0.12$, and $J=0.4$ for electron doping. The real value of $t$ in cuprates is usually taken to be 0.35eV. The dynamical charge correlation function is defined as

$$N(\mathbf{q},\omega) = \frac{1}{\pi} \text{Im} \langle 0 | N_{-\mathbf{q}} \frac{1}{H - E_0 - \omega - i\gamma} N_{\mathbf{q}} | 0 \rangle,$$

where $|0\rangle$ is the ground state with energy $E_0$ and $N_{\mathbf{q}} = N^{-1/2} \sum_{i,\sigma} e^{i\mathbf{q}\cdot\mathbf{R}_i} \tilde{c}^\dagger_{i,\sigma} \tilde{c}_{i,\sigma}$ with number of site $N$ and the position vector $\mathbf{R}_i$ at site $i$. The small positive number $\gamma$ is set to be 0.02 in our calculation. The doping dependence of $N(\mathbf{q},\omega)$ for a $N = 20$ system is shown in Fig. S2, where x = 0.1 (x = 0.2) corresponds to the case of 2 (4) electrons in the system. An averaging procedure for twisted boundary conditions is used to reduce the finite-size effect [31].

**Supplementary reference**

[31] T. Tohyama, Asymmetry of the electronic states in hole- and electron-doped cuprates: Exact diagonalization study of the *t-t'-t''-J* model, *Phys. Rev. B* **70**, 174517 (2004).

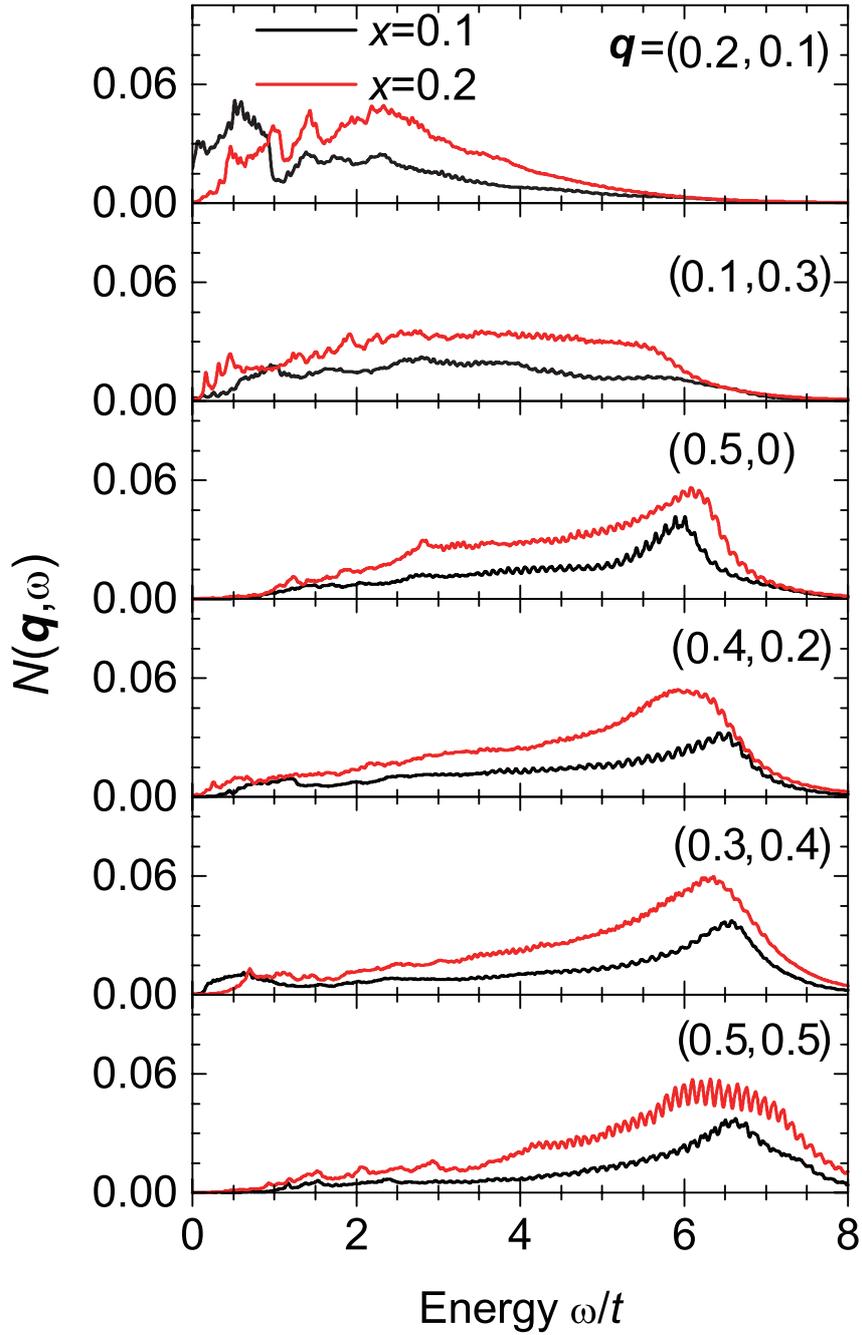

**Figure S2 | Dynamical charge correlation function $N(q,\omega)$ for an electron-doped $N=20$ $t$-$t'$-$t''$-$J$ lattice with three-site terms, obtained by averaging over twisted boundary condition.** Black (red) curve corresponds to the carrier concentration of x=0.1 (x=0.2).